# Strong electron-phonon coupling in δ-phase stabilized Pu

M. J. Graf, T. Lookman, J. M. Wills, D. C. Wallace and J. C. Lashley

*Los Alamos National Laboratory, Los Alamos, NM 87545, USA*

(Dated: May 25, 2005- LA-UR-05-0904)

Heat capacity measurements of the δ-phase stabilized alloy $Pu_{0.95}Al_{0.05}$ suggest that strong electron-phonon coupling is required to explain the moderate renormalization of the electronic density of states near the Fermi energy. We calculate the heat capacity contributions from the lattice and electronic degrees of freedom as well as from the electron-lattice coupling term and find good overall agreement between experiment and theory assuming a dimensionless electron-phonon coupling parameter of order unity, $\lambda \sim 0.8$. This large electron-phonon coupling parameter is comparable to reported values in other superconducting metals with face-centered cubic crystal structure, for example, Pd ($\lambda \approx 0.7$) and Pb ($\lambda \approx 1.5$). Further, our analysis shows evidence of a sizable residual low-temperature entropy contribution, $S_{res} \approx 0.4\,k_B$ (per atom). We can fit the residual specific heat to a two-level system. Therefore, we speculate that the observed residual entropy originates from crystal-electric field effects of the Pu atoms or from self-irradiation induced defects frozen in at low temperatures.

## I. INTRODUCTION

Plutonium, a member of the actinides exhibits six unique crystal structures (phases) in the solid state at ambient pressure between absolute zero and its melting temperature. The phases range in symmetry from simple monoclinic (sm) to body-centered cubic. One phase, the easily worked face-centered cubic (fcc) phase, denoted by δ, is thermodynamically stable in pure plutonium from 592 K to 736 K, and can be stabilized down to room temperature by small additions of trivalent elements such as gallium or aluminum. The effects of alloying are not only structural but have a profound impact on the electronic structure. Specifically, the low-temperature α phase (sm) has a very large coefficient of thermal expansion and a Sommerfeld coefficient of $\gamma_S \approx 17\,\mathrm{mJ}/(\mathrm{mol\,K^2})$, while the high-temperature δ phase exhibits a moderately enhanced Sommerfeld coefficient $\gamma_S \approx 50-70\,\mathrm{mJ}/(\mathrm{mol\,K^2})$, depending on concentration and negative coefficient of thermal expansion.[1–7] Change in structure, the electronic specific heat, and the sign of thermal expansivity with a modest amount of trivalent atoms in the δ phase are not fully described by electronic structure theory. It is believed that the high-temperature and high-volume phase has localized, nonbinding electrons, while in the low-temperature and low-volume phase the electrons are itinerant and binding. This behavior is similar to the Mott transition in correlated electron systems. Some of this behavior is borne out in recent electronic structure calculations.[8–10] In a recent study, using inelastic neutron scattering McQueeney and coworkers[11] reported unusual softening of phonons and elastic moduli in $Pu_{0.95}Al_{0.05}$. Using the same sample, Lashley and coworkers[12] pointed out that the low-temperature data of the heat capacity exhibits a moderately enhanced Sommerfeld coefficient, $\gamma_S \approx 64\,\mathrm{mJ}/(\mathrm{mol\,K^2})$, and a λ-shaped anomaly in the heat capacity around 60 K. These observations were suggestive of describing $Pu_{0.95}Al_{0.05}$ as an incipient heavy-fermion system.

The purpose of this study is to give a quantitative description of the electron-phonon interaction on the conduction electrons in plutonium, and whether the observed low-temperature anomaly in the heat capacity is associated with a martensitic phase transformation. In this paper, we present calculations of the temperature dependence of the mass enhancement of the conduction electrons which agree well with the magnitude of the observed specific heat data in $Pu_{0.95}Al_{0.05}$, enriched with 95% of the isotope $^{242}$Pu. A remaining residual entropy contribution, after subtraction of electron, phonon, and electron-phonon terms from the measured data, is unlikely due to a partial martensitic phase transformation from the high-temperature δ phase into the low-temperature $\alpha'$ phase.[11,12] On cooling this transformation finishes around 130 - 180 K and is completely reversed on heating around 380 K.[13,14] Instead of a structural transformation, we speculate that crystal-electric field effects or self-irradiation induced defects and vacancies, for example, Frenkel pairs, are responsible for the excess entropy. Additionally, the relatively strong electron-phonon coupling of order unity, necessary for describing the measured specific heat data, would suggest that $Pu_{0.95}Al_{0.05}$ should become superconducting below a few Kelvin. So far no evidence of superconductivity has been observed down to roughly 3 K.

## II. THEORY

We follow the standard approach and divide the calculation of the total heat capacity of a metal into a vibrational, electronic, electron-phonon coupling, and residual (everything else) term, $C = C_{ph} + C_e + C_{ep} + C_{res}$. One by one, we will calculate the contribution of each term and determine its importance. Since thermal expansion of $Pu_{0.95}Al_{0.05}$ is negligible over the entire temperature range of interest, i.e., $1 \leq C_P/C_V < 1.006$, we will assume in our analysis that $C_P \simeq C_V$ and drop the subscript. Furthermore, we will assume in our study that the thermodynamic properties are dominated by the fcc δ-Pu crystal structure and any possible admixture of $\alpha'$-Pu is negligible. This assumption is justified by previous resistivity,[14] neutron diffraction,[15] inelastic neutron scattering and ultrasound measurements[11] on $Pu_{0.95}Al_{0.05}$. They show either no structural transformation between 4 K and 300 K, or indicate on cooling a partial transformation of at most 3-5% of the sample. The partial transformation from δ to α′ occurs around 130 - 180 K and is possibly due to metastable surface formation during sample preparation.[11]

The effects of electron-phonon interaction in metals has been studied extensively in the past and is well understood within the framework of the strong coupling theory of Eliashberg.[16,17] We follow the approach by Grimvall,[18,19] Bergman et al.,[20] and Allen.[21] In the presence of the electron-phonon interaction the electronic quasiparticle spectrum is modified by

$$E(\mathbf{k}) = \xi_\mathbf{k} + \mathrm{Re}\,\Sigma_{ep}(\mathbf{k}, E(\mathbf{k}); T), \quad (1)$$

where $\xi_\mathbf{k} = \varepsilon_\mathbf{k} - \mu$, $\mu$ is the chemical potential, and the shift due to the electron-phonon self-energy $\Sigma_{ep}$ yields the electron mass enhancement. The finite temperature electron-phonon self-energy is[18,19]

$$\begin{aligned}\Sigma_{ep}(\mathbf{k},\omega) &= \frac{1}{V}\sum_s\sum_{\mathbf{k'}}|g_s(\mathbf{q})|^2 \\ &\times \left(\frac{1-f_{\mathbf{k'}}+b_{s\mathbf{q}}}{\omega-\xi_{\mathbf{k'}}-\omega_{s\mathbf{q}}}+\frac{f_{\mathbf{k'}}+b_{s\mathbf{q}}}{\omega-\xi_{\mathbf{k'}}+\omega_{s\mathbf{q}}}\right),\end{aligned} \quad (2)$$

where $V$ is the volume, and $f_{\mathbf{k'}}$ and $b_{s\mathbf{q}}$ are the usual Fermi-Dirac and Bose-Einstein factors. The phonon frequencies $\omega_{s\mathbf{q}}$ are labeled by branch index $s$ and momentum $\mathbf{q}$. The strength of the electron-phonon coupling for momentum transfer $\mathbf{q} = \mathbf{k} - \mathbf{k'}$ is denoted by $g_s(\mathbf{q})$. Integration over the Fermi surface and averaging over all directions $\mathbf{k}$ gives for an isotropic electron-phonon interaction[18,19]

$$\begin{aligned}\Sigma_{ep}(\omega) &= \int d\varepsilon [N(\varepsilon)/N(E_F)] \int d\omega' \alpha^2 F(\omega') \\ &\times \left(\frac{1-f(\varepsilon)+b(\omega')}{\omega-\varepsilon+\mu-\omega'}+\frac{f(\varepsilon)+b(\omega')}{\omega-\varepsilon+\mu+\omega'}\right),\end{aligned} \quad (3)$$

where $N(\varepsilon)$ is the electron density of states and the Eliashberg function $\alpha^2 F(\omega)$ is the usual product of the electron-phonon interaction with the phonon density of states and the electron density of states at the Fermi energy $E_F$.

### A. Phonon density of states

The lattice heat capacity of a crystal with quasi-harmonic vibrations is determined by knowing the phonon dispersion or equivalently its phonon density of states (PDOS). We obtain the PDOS of $Pu_{0.95}Al_{0.05}$ by fitting a Born-von Kármán force model, including up to three next-neighbor (3NN) atomic shells, to phonon dispersions at room temperature on δ-phase stabilized Pu alloyed with Ga, recently measured by Wong and coworkers.[22] To avoid potentially small differences in the low-temperature properties of the calculated heat capacity between Pu alloyed with Ga and Al, we simultaneously fitted the dispersion curves for $Pu_{0.98}Ga_{0.02}$ and the elastic moduli for $Pu_{0.95}Al_{0.05}$[11] at room temperature. The fitted moduli are $C_{11} = 34.0\,\mathrm{GPa}$, $C_{44} = 31.4\,\mathrm{GPa}$, and $C_{12} = 24.9\,\mathrm{GPa}$. In the remainder of this work this will be crucial for a proper analysis of the heat capacity.

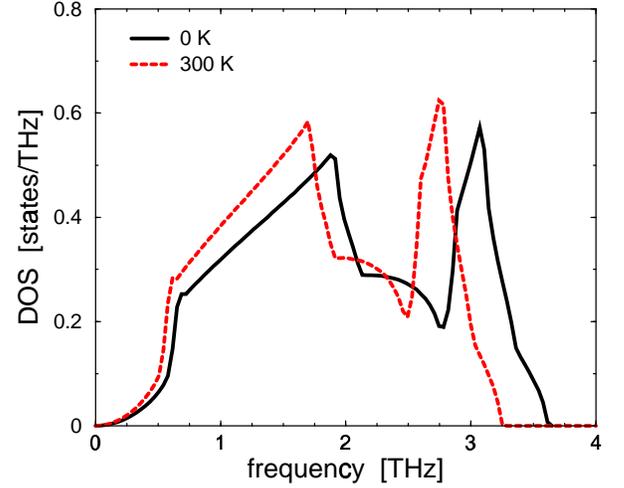

FIG. 1: (Color online) Phonon density of states (PDOS) calculated by fitting a 3NN Born-von Kármán force model to measured phonon dispersions and elastic coefficients at room temperature.[11,22] Extrapolation to lower temperatures is obtained by renormalizing all interatomic force constants with the measured elastic coefficients.[11]

In Fig. 1 we show the corresponding phonon density of states at 300 K and at absolute 0 K computed for $Pu_{0.95}Al_{0.05}$. The noticeable difference in the high and low temperature PDOS is due to the large temperature dependence of the elastic coefficients, which vary by over 20% over this temperature range. Since no low-temperature phonon dispersions are available, we rescaled all interatomic force constants at every temperature by the measured elastic coefficients, $C_{ij}(T)/C_{ij}(300\,\mathrm{K}) \approx 1.24 - 8 \cdot 10^{-4}T/\mathrm{K}$, and recalculated the phonon dispersions and PDOS. This procedure guarantees the correct low and high temperature long wavelength phonon dispersions. Note that the elastic bulk and shear moduli were measured on a polycrystalline sample, thus no informa-tion about the shear anisotropy $2C_{44}/(C_{11}-C_{12})$ is available. Therefore, in our calculations we assumed that $Pu_{0.95}Al_{0.05}$ has the same anomalous shear anisotropy ($\sim 7$) over the entire temperature range as its sister alloy $Pu_{0.98}Ga_{0.02}$ at room temperature.[23] This assumption is justified by the negligible thermal expansion and similar temperature behavior of the polycrystalline bulk and shear moduli of $Pu_{0.95}Al_{0.05}$.

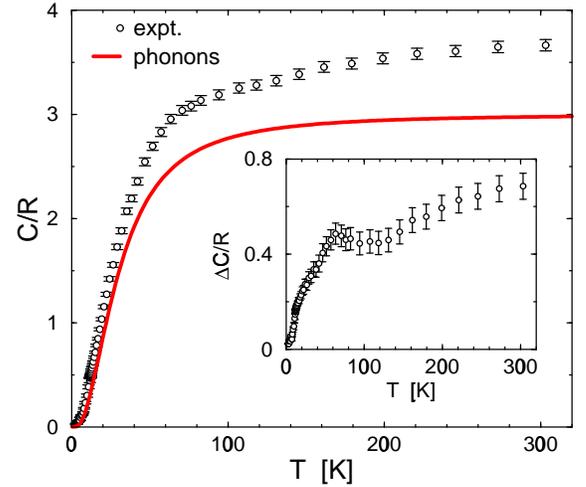

FIG. 2: (Color online) Measured heat capacity $C$ ($C_P \simeq C_V$) of $Pu_{0.95}Al_{0.05}$ and calculated vibrational contribution $C_{ph}$. Inset: Difference curve $\Delta C$.

The calculated specific heat of the quasi-harmonic crystal

$$C_{ph} = \frac{1}{V}\sum_{s\mathbf{q}} \hbar\omega_{s\mathbf{q}} \frac{db_{s\mathbf{q}}}{dT} \qquad (4)$$

saturates at high temperatures at the Dulong-Petit limit of $3R$ for a classical phonon gas, where $R=Nk_B$ is the gas constant per mole, and accounts for most of the experimental data seen in Fig. 2. The difference curve $\Delta C = C - C_{ph}$ shown in the inset of Fig. 2, which is similar to the one in Fig. 1 of Ref. 12, will be discussed shortly in more detail.

Knowing the PDOS of a system, we can now compute any phonon moment of interest. The most important moments for thermodynamic studies are the logarithmic moment $\omega_0$, and $\omega_1$, $\omega_2$, and $\omega_{-3}$. The latter one determines the Debye temperature $k_B\Theta_D = \hbar\omega_{-3}$. The log-moment $\omega_0$ and the quadratic moment $\omega_2$ enter the high-temperature expansions of thermal functions and therefore are calculated at room temperature. The linear moment $\omega_1$ measures the zero-point energy vibrations and is calculated for comparison at room temperature, too. We computed all moments directly (see Table I) except for the Debye temperature. $\Theta_D$ was extracted from the calculated low-temperature heat capacity in Fig. 3 and is in very good agreement with experiment. Until now these are the most accurately determined phonon moments of $\delta$-phase stabilized Pu and generally differ by 10-20% from published values in the literature. If we extrapolate the temperature dependence of the measured elastic moduli to 600 K and recalculate the phonon moments, then we can compare directly with Wallace's estimates for the pure $\delta$ phase of plutonium. At 600 K we calculate $\Theta_0 = 68\,\text{K}$ and $\Theta_2 = 104\,\text{K}$, which is in very good agreement with Wallace's log-moment. Note that he used a constrained analysis for the entropy, assuming $\Theta_2 \simeq e^{\frac{1}{3}}\Theta_0$, because the high-temperature behavior of the entropy is dominated by the log-moment. The phonon moments by McQueeney et al.[11] were obtained from inelastic neutron scattering on a polycrystalline sample and are consistently 8% bigger than ours. This discrepancy is (1) due to a higher maximum phonon frequency in their PDOS, the origin of which is not yet understood, and (2) probably due to a misprint of the values of the longitudinal speed of sound in their paper. The Debye temperature reported earlier by Lashley et al.[12] on the same sample is $\sim 10\%$ smaller than ours, because in their data analysis the Debye approximation was applied outside its validity region, which is $T \lesssim \Theta_D/50 \approx 2.3\,\text{K}$, and which is just below their lowest data point, as can be seen in Fig. 3.

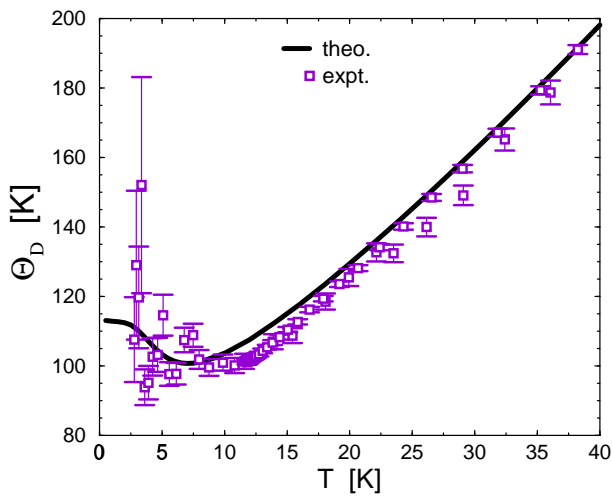

FIG. 3: (Color online) Debye temperature extracted from the low-temperature approximation $C_{ph}/R = \frac{12\pi^4}{5}(T/\Theta_D)^3$. A linear temperature term was subtracted from the experimental $C$. The large error bars at low $T$ are a consequence of the subtraction procedure.

TABLE I: Phonon moments of $Pu_{0.95}Al_{0.05}$ calculated at 300 K, except for the Debye temperature $\Theta_D$ which is calculated at $T \to 0$. We use the conversion $k_B\Theta_n = \hbar\omega_n$ and $k_B\Theta_0 = e^{-\frac{1}{3}}\hbar\omega_0$. Note that Wallace[24] reported moments for pure $\delta$-Pu near 600 K and assumed $\Theta_2 \simeq e^{\frac{1}{3}}\Theta_0$. The estimated error of the last digit is given in parentheses.

|           | $\Theta_D$ / K | $\Theta_0$ / K | $\Theta_1$ / K | $\Theta_2$/K |
|-----------|---------|---------|---------|---------|
| this work | 116(2)  | 77(2)   | 113(2)  | 118(2)  |
| McQueeney | 125     | 84      | 122     | 127     |
| Wallace   |         | 66      |         | 93      |
| Lashley   | 100(2)  |         |         |         |

### B. Electron density of states

In a normal metal, the temperature dependence of the specific heat at low temperatures is dominated by the electronic term, which is $C_e = \gamma_0 T$ for a nearly-free electron gas with a flat electron density of states (EDOS) in the vicinity $k_BT$ of the Fermi energy $E_F$ with Sommerfeld constant $\gamma_0$. If the EDOS is structured and peaked near $E_F$, then the $T$-dependence of $C_e$ is more complicated and needs to be calculated from

$$C_e = \frac{2}{V}\sum_\mathbf{k} \xi_\mathbf{k} \frac{df(\xi_\mathbf{k})}{dT} . \qquad (5)$$

Integration over the Fermi surface yields

$$C_e = 2\int d\xi N(\xi)\xi \frac{df(\xi)}{dT} . \qquad (6)$$

First-principles electronic structure calculations for fcc $\delta$-Pu show a peaked behavior of the $5f$ electrons near $E_F$ in the density of states, in good agreement with photoemission measurements.[25] Since first-principles calculations do not have the energy resolution of a few milli electron-volts, we model the EDOS of $Pu_{0.95}Al_{0.05}$ by three Lorentzians with different constant backgrounds above and below $E_F$ (see Fig. 4).

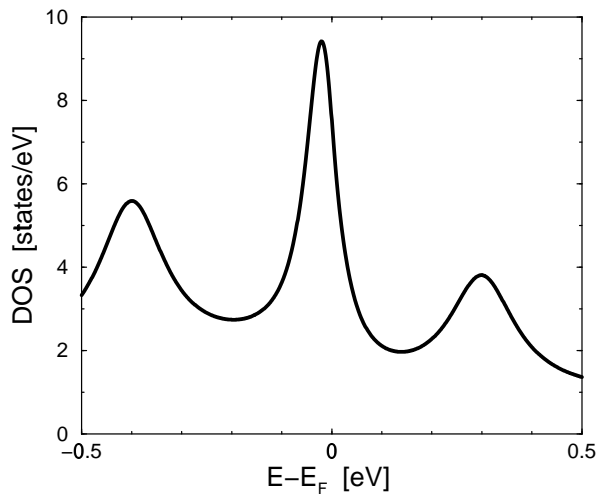

FIG. 4: (Color online) Model for the one-electron density of states (EDOS) per unit cell of $Pu_{0.95}Al_{0.05}$ in the vicinity of the Fermi energy $E_F$.

Remarkably, the temperature dependence of the heat capacity at high and low temperatures strongly constrains the possible shape of the EDOS models. We found after several adjustments that an EDOS model with three Lorentzians located at energies $E - E_F = -0.4, -0.02, 0.3$ eV, and differently scaled constant backgrounds above and below $E_F$, reproduces $C/T$ very well at high temperatures, $T > 100$ K, where electron-phonon coupling is negligible, and requires only a modest electron-phonon parameter $\lambda$ at zero temperature. We did not attempt to further refine this EDOS model, nor did we account for a weak temperature dependence of the chemical potential in our calculations, due to the particle-hole asymmetry of the resonance near the Fermi surface. We set the chemical potential equal to the Fermi energy in the temperature range from $0 - 300$ K. For now, we neglect such higher order effects in our crude approximation of the electronic heat capacity and the electron-phonon self-energy. Because this also touches on the difficult problem of how to accurately treat the occupation of localized versus itinerant $5f$ electrons in this material, which is still an unsolved problem. However, we found numerically that the electron-phonon self-energy is very insensitive to the structured EDOS model, in contrast to the electronic heat capacity.

### C. Electron-phonon renormalization of heat capacity

We apply Migdal's approximation to the calculation of the electron-phonon self-energy,[26] as outlined earlier, and obtain the standard results for the renormalization of the EDOS or equivalently for the mass of the conduction electrons. The electron-phonon interaction renormalizes the electronic specific heat according to $C_e \to C_e + C_{ep}$:[18–20,27]

$$C_e + C_{ep} = \frac{2}{V} \sum_{\mathbf{k}} E_{\mathbf{k}} \frac{df(E_{\mathbf{k}})}{dT}$$

$$= 2 \int dE\, N(E) E \left[ [1 - \partial_E \text{Re} \Sigma_{ep}(E)] \partial_T f(E) \right.$$

$$\left. + \partial_T \text{Re} \Sigma_{ep}(E) \partial_E f(E) \right] . \quad (7)$$

This is conveniently written as

$$C_e + C_{ep} = \left[ \gamma_0(T) + \gamma_{ep}(T) \right] T . \quad (8)$$

For a flat EDOS the bare electronic heat capacity of a metal is $C_e = \gamma_0 T$, but more generally $\gamma_0 \to \gamma_0(T)$. The electron-phonon term is related to the electron-phonon parameter $\lambda$ at $T = 0$ and the unnormalized Sommerfeld coefficient $\gamma_0$ by

$$\gamma_{ep}(T) = \lambda \gamma_0(0) \left( \gamma_{ep}(T)/\gamma_{ep}(0) \right) . \quad (9)$$

This standard calculation neglects any effects of anharmonic or nonadiabatic phonons on the vibrational heat capacity and the electron-phonon self-energy that go beyond the quasiharmonic approximation. Various theories have been developed that go beyond Migdal's approximation.[28–31] But our analysis of the heat capacity of $Pu_{0.95}Al_{0.05}$ (see Figs. 8 and 11, and the discussion at the end of this section) shows no significant contribution of anharmonic phonons up to room temperature, despite a low Debye temperature. For example, the Debye temperature of Pb is $\Theta_D = 105$ K. Thus, we discard any anharmonic or nonadiabatic corrections to the calculation of the heat capacity.

Since we have no knowledge of the frequency dependence of Eliashberg's function $\alpha^2 F(\omega)$ for $Pu_{0.95}Al_{0.05}$, we will use (1) a very simplistic model with a single Einstein oscillator that describes the coupling between the conduction electrons and the lattice near the longitudinal zone boundary phonons, and (2) a more realistic fcc spectrum model using the PDOS, where each phonon mode couples equally to the conduction electrons. The *true* $\alpha^2 F(\omega)$ will lie somewhere between the

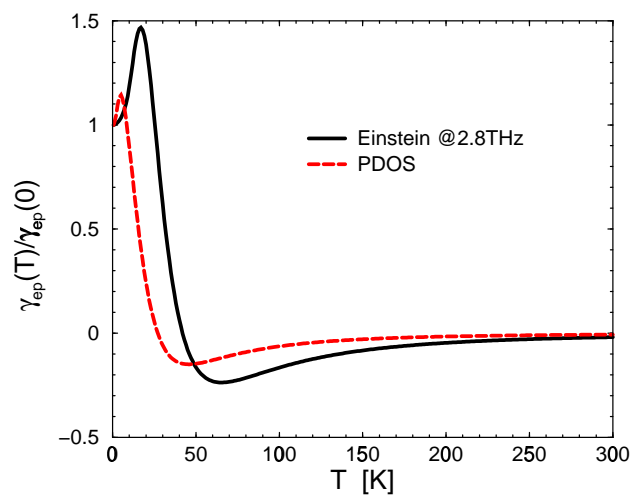

FIG. 5: (Color online) Electron-phonon enhancement of the electronic coefficient in the heat capacity assuming that (a) the Eliashberg function has a dominant Einstein mode at 2.8 THz, $\alpha^2 F(\omega) \propto \delta(\omega - \omega_E)$, and (b) $\alpha^2 F(\omega)$ is proportional to the PDOS.

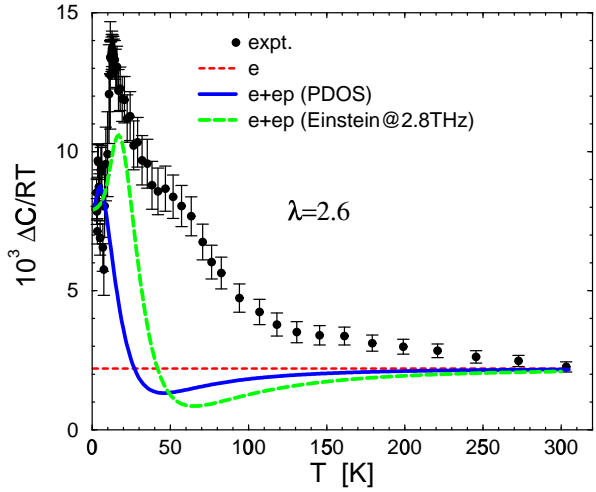

FIG. 6: (Color online) Electron-phonon enhancement of the electronic heat capacity assuming a free electron gas, i.e., constant EDOS near $E_F$ combined with an $\alpha^2 F$-function that has either an Einstein mode or a realistic fcc spectrum (PDOS shown in Fig. 1). For comparison the difference curve after subtraction of the lattice contribution, $\Delta C/T = (C - C_{ph})/T$ (solid circles) and the bare electronic heat capacity $C_e/T$ (e: dashed line) are shown.

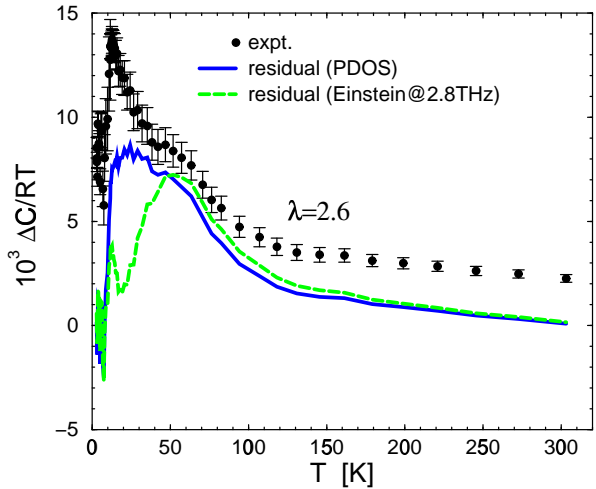

FIG. 7: (Color online) Residual heat capacity after subtracting electron (flat EDOS), phonon and electron-phonon contributions from measured $C/T$. For comparison $\Delta C/T$ is shown (solid circles).

Einstein model and the fcc spectrum model and result in an electron-phonon enhancement of the electronic heat capacity that is approximated by the limits shown in Fig. 5.

For simplicity we kept the chemical potential fixed at the Fermi level for all temperatures when computing the electron-phonon self-energy. Because this affects the electronic heat capacity more than the electron-phonon self-energy, as discussed before, we neglected any corrections to $\alpha^2 F(\omega)$ arising from a shift of the chemical potential. For two dimensions Mahan[32] discussed some of the aspects of electron-phonon interaction near Van Hove singularities in the context of high-temperature superconductors. However, this is a complex and difficult problem and goes far beyond the scope this analysis.

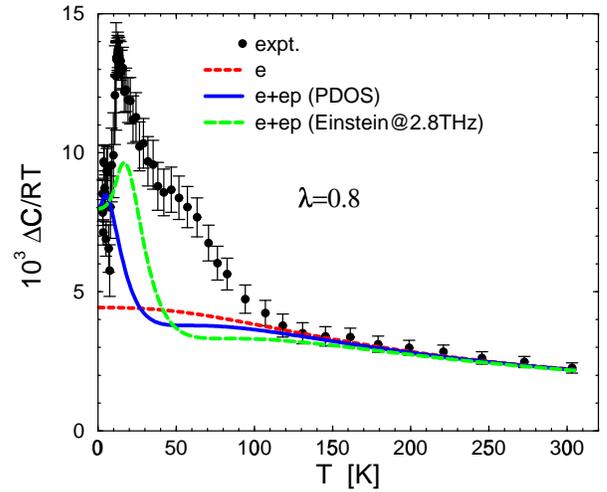

FIG. 8: (Color online) Electron-phonon enhancement of the electronic heat capacity using the EDOS model from Fig. 4 combined with an $\alpha^2 F$-function that has either an Einstein mode or a realistic fcc spectrum (PDOS from Fig. 1). For comparison $\Delta C/T$ (solid circles) and the bare electronic $C_e/T$ (e: dashed line) are shown.

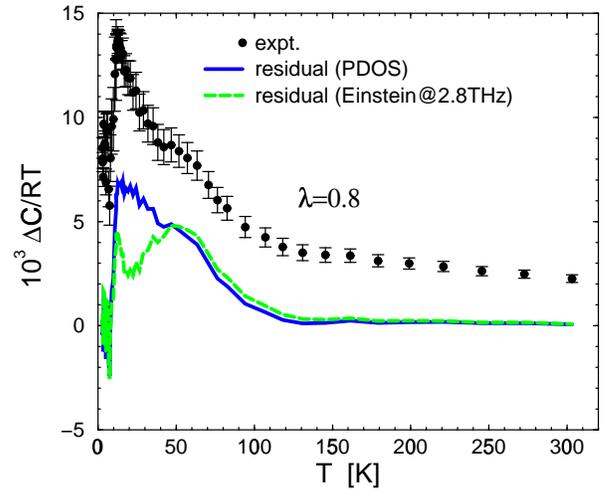

FIG. 9: (Color online) Residual heat capacity after subtracting electron (peaked EDOS), phonon and electron-phonon contributions from measured $C/T$. For comparison $\Delta C/T$ is shown (solid circles).

We can give an upper estimate of the dimensionless electron-phonon coupling constant $\lambda$ by assuming a flat EDOS. This means that all of the low-temperature enhancement of the Sommerfeld coefficient relative to its high-temperature value is due to the electron-phonon interaction. However, it neglects electronic band structure or many-electron effects. In Fig. 6 we report calculations for the bare and electron-phonon renormalized electronic heat capacity. The latter requires an unrealistically large electron-phonon coupling parameter $\lambda = 2.6$. The Einstein oscillator model gives a slightly better account of the low-temperature behav-ior than the fcc spectrum model. The remaining discrepancy between the difference curve $\Delta C$ and theory is shown in Fig. 7. The residual entropy between the integrated curves $\Delta C/T$ and $(C_e + C_{ep})/T$ is unexpectedly big and roughly $S(303K) = R \int_{0K}^{303K} dT [\Delta C - C_e - C_{ep}]/T \approx 0.7R$.

Next we give a lower estimate for $\lambda$ by assuming an EDOS peaked near the Fermi level. This time we include all band structure and electronic correlation effects into our model EDOS (Fig. 4). In Fig. 8 we report the bare and electron-phonon renormalized electronic heat capacity. We estimate the coupling parameter $\lambda = 0.8$, which is in remarkably good agreement with the thermodynamic estimate by Wallace ($\lambda = 0.85$).[24] Again the Einstein model for $\alpha^2 F$ gives a slightly better account of the low-temperature behavior than the realistic spectrum model. The remaining discrepancy between the difference curve $\Delta C$ and theory is shown in Fig. 9. The residual entropy between the integrated curves $\Delta C/T$ and $(C_e + C_{ep})/T$ is somewhat smaller and roughly $S(303K) = R \int_{0K}^{303K} dT [\Delta C - C_e - C_{ep}]/T \approx 0.4R$, with negligible entropic contribution above $\sim 100$ K. Note the extremely small bump in the residual heat capacity around $T_M \approx 160$ K in Figs. 9 and 10, which is within the experimental and theoretical uncertainties. The residual entropy associated with this bump between 118 K and 303 K is indeed very small, $\Delta S_M = S(303K) - S(118K) \approx 0.028 - 0.043R$.

Our next concern is whether or not the residual specific heat can be attributed to lattice anharmonicity. Indeed, this is not the case, because the curve in Fig. 10 is uncharacteristic of anharmonicity both in temperature dependence and in magnitude.[33] First, anharmonicity is never found to have a significant contribution only within a narrow temperature range at low temperatures. Second, the estimated residual entropy of $0.4 - 0.7R$ at 303 K is much too large for anharmonicity, since the anharmonic entropy $S_{anh}$ at 303 K is less than $0.1R$ for all the analyzed elemental metals.[33,34]

The key result of this work is that after accounting for the heat capacities from phonons, electrons, and electron-phonon interaction there is still excess entropy remaining that awaits explanation.

## III. STRUCTURAL PHASE TRANSFORMATION $\delta \to \alpha'$

Our analysis of the heat capacity has shown that there is an unaccounted for residual excess entropy of roughly $S_{res} \sim 0.4R - 0.7R$ located between $10 \text{K} - 100 \text{K}$. This low-temperature behavior is clearly separated from the "high-energy physics" of 1400 K that follows from the invar model proposed by Lawson and coworkers for the $\delta$-phase stabilized alloys of plutonium.[35,36] Therefore, it is tempting to attribute this excess entropy to a partial martensitic-like phase transformation.[12] There is ample experimental evidence that on cooling parts of the $\delta$-phase stabilized Pu undergo a martensitic transition from the $\delta$ phase to the $\alpha'$ phase.[14,37,38] The $\alpha'$ phase has the same space group as the monoclinic $\alpha$ phase, except that some of the Pu atoms have been substituted by the alloying element.

If the structural transformation occurs at temperatures above $\Theta_0$, i.e., in the classical high-temperature limit, then the estimated vibrational entropy difference is

$$\Delta S_{ph} = S_{ph}^\delta - S_{ph}^\alpha \approx 3R \ln(\Theta_0^\alpha / \Theta_0^\delta). \quad (10)$$

Taking $\Theta_0^\delta = 77$ K and $\Theta_0^\alpha = 116$ K (from Ref. 33), we find $\Delta S_{ph} \approx 1.23R$. This agrees with the total entropy increase across three phase transitions, as measured experimentally at ambient pressure in pure plutonium and corrected for the electronic contribution, $\Delta S_{ph}^{\alpha\delta} = \Delta S_{ph}^{\alpha\beta} + \Delta S_{ph}^{\beta\gamma} + \Delta S_{ph}^{\gamma\delta} \approx 1.4R$.[24,39–41] However, since only 3-5% of the sample transforms between $\delta$ and $\alpha'$, the available transformational entropy is around $0.04 - 0.07R$, far too small to account for $\Delta S_{res} \approx 0.4 - 0.7R$. Further, as mentioned in Sec. II, no transformation in either direction takes place below $\sim 130$ K. Therefore, the potential martensitic transformation cannot be the cause for the residual specific heat, which appears at still lower temperatures, see Fig. 10.

Alternatively, we can estimate the amount of latent heat or entropy difference by applying Landau's theory of phase transitions to the martensitic $\delta$ to $\alpha'$ transition. Various martensitic transformation paths between cubic and monoclinic symmetries have been discussed for the alloys Ni-Ti and Cu-Zn-Al.[43] Here, the situation is slightly different as discussed by Mettout et al.,[44] because the axis of monoclinic (two-fold) symmetry of the martensite is along the $b$ axis ($\langle 010 \rangle$ direction) of the cubic austenite. Therefore, the simplest elastic free energy per volume $V$, describing a transformation at temperature $T_M$, is[42]

$$\Delta F = (T - T_M) \left( \partial_T C_{44} \varepsilon_{xy}^2 + \frac{1}{2} \partial_T B \varepsilon_B^2 \right), \quad (11)$$

with $\varepsilon_B^2 \equiv \varepsilon_{xx}^2 + \varepsilon_{yy}^2 + \varepsilon_{zz}^2$. Its corresponding entropy difference, $\Delta S = -\partial_T \Delta F$, is

$$\Delta S = -\partial_T C_{44} \varepsilon_{xy}^2 - \frac{1}{2} \partial_T B \varepsilon_B^2 \approx 3.14 \cdot 10^{-2} \frac{\text{GPa}}{\text{K}} \cdot 2.20 \cdot 10^{-3} + 1.27 \cdot 10^{-2} \frac{\text{GPa}}{\text{K}} \cdot 3.55 \cdot 10^{-2}, \quad (12)$$

or per mole of plutonium atoms we have $\Delta S \approx (0.12 + 0.79)R$. The strain order parameters $\varepsilon_{xy} = 0.0469$, $\text{tr}\,\varepsilon = -0.1621$, and $\varepsilon_B^2 = 0.0355$, were calculated from published lattice parameters of the high- and low-symmetry phases at 140 K, with $a_0 = 4.5959$ Å and $a = 6.1328$ Å, $b = 4.7824$ Å, $c = 10.8997$ Å, $\beta = 101.816°$, respectively.[15,45,46] Note that $\varepsilon_{xy}$ and $\text{tr}\,\varepsilon$ are independent of the angle $\beta$. The temperature derivatives of the shear modulus, $G$, and bulk modulus, $B$, have been reported for the high-symmetry (austenite) $\delta$ phase.[11] However, lacking measurements of the single-crystal elastic shear modulus $C_{44}$, we used the measured polycrystalline average assuming $\partial_T C_{44} \sim 2 \partial_T G$, because of (1) $C_{44} \approx 2G$ at room temperature and (2) the similar temperature dependence of $G$ and $B$.[11] The estimate for $\Delta S$, based on continuum theory for deformations, is in reasonably good agreement with the high-temperature estimate (1.2R), which is based on lattice dynamic theory, and the measured entropy difference across three phases in pure plutonium ($1.4R$). Thus the predicted elastic entropy difference of roughly $0.9R$ is consistent with a direct transformation path between the cubic and monoclinic symmetries in $Pu_{0.95}Al_{0.05}$, but does not explain the excess entropy at low temperatures.

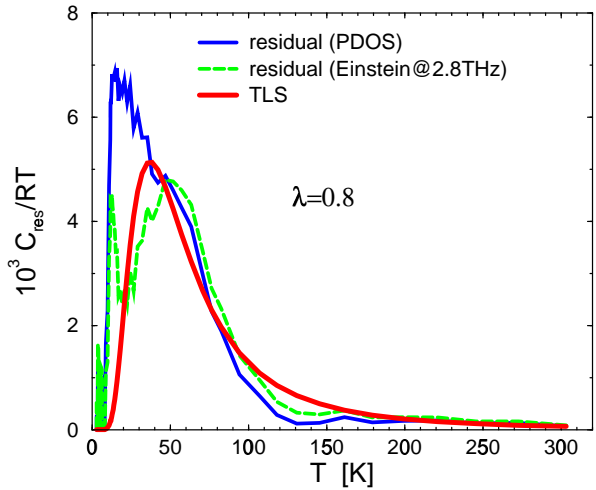

FIG. 10: (Color online) Heat capacity over temperature ($C/RT$) of a two-level system with energy separation $T_{TLS} \approx 120\,\mathrm{K}$ and site occupation $n \approx 0.5$; residual heat capacities are from Fig. 9.

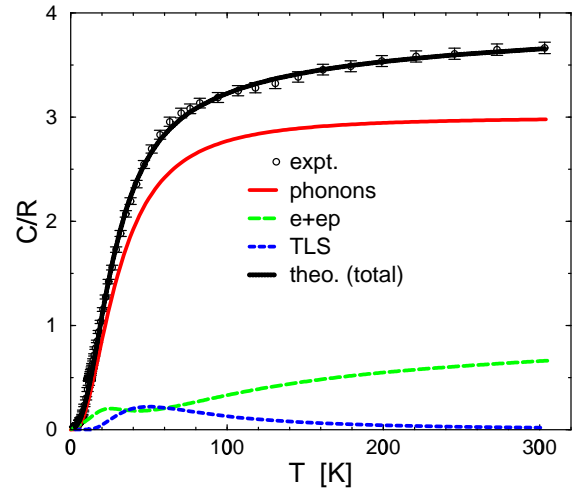

FIG. 11: (Color online) Experimental and combined total theoretical heat capacities.

So far we have not considered any electronic change in entropy. For a full description, we need to address the amount of electronic entropy difference between the δ and α′ phases. If we use the peaked EDOS model to compute $C_e/T$ for the δ phase (see Fig. 8) and subtract the high-temperature Sommerfeld coefficient, as a best estimate for the electronic contribution of the α′ phase, then we find $\Delta S_e(100K) \approx 0.2R$ and $\Delta S_e(303K) \approx 0.34R$. This suggests that the transformational entropy difference is dominated by the contribution from lattice vibrations, justifying our phonon analysis.

Thus the proposed scenario[12] of a martensitic phase transformation, occurring between 130 K and 380 K, is unlikely to explain the excess entropy or the λ-shaped heat capacity below 100 K. Since at most 3-5% of the sample transforms, one would expect to observe an entropy difference of $\Delta S \lesssim 0.04 - 0.07R$ between 130 K and 380 K. Indeed, this is consistent with the observed entropy difference $\Delta S_M \approx 0.028 - 0.043R$ between 118 K and 303 K. However, this is much smaller than the low-temperature excess entropy $\Delta S_{res} \approx 0.4 - 0.7R$, which occurs mostly below 110 K, as can be seen in Fig. 10.

## IV. HEAT CAPACITY OF LOCALIZED $5f$ ELECTRONS

Of course there are other possibilities for additional degrees of freedom to explain an excess in entropy, for example, a magnetic contribution due to localized magnetic moments. However, very recently Lashley and coworkers[47] analyzed various experiments and argued vehemently against any form of local magnetic moments in α and δ phase plutonium at low temperatures. Alternatively, crystal-electric field ef-fects may play a significant role in plutonium and its alloys.[48] It has been argued for the system of Pu monopnictides that the ground state is a nonmagnetic singlet and the first excited crystal-electric field level is on the order of 100 K above the ground state.[49,50] Therefore, each $5f$ electron site will show an additional internal degree of freedom that contributes to the total entropy. For simplicity, we model the localized $5f$ electrons as independent two-level systems (each Pu atom has two crystal-electric field levels with net-zero spin), then we obtain a good fit to $S_{res}$ by calculating

$$C_{TLS}/R = n \left(\frac{T_{TLS}}{T}\right)^2 \frac{\exp(T_{TLS}/T)}{[1+\exp(T_{TLS}/T)]^2},\qquad(13)$$

with fit parameters for site occupation $n \approx 0.5$ and level splitting $T_{TLS} \approx 120\,\mathrm{K}$ (see Fig. 10). However, it is unsatisfactory that only every other plutonium atom ($n \approx 0.5$) should contribute. On the other hand, one may argue that this agrees with the scenario by Cooper and coworkers,[51,52] who have argued that a fraction of Pu atoms has fluctuating valences, $Pu^{4+/5+}$, due to the presence of Ga or Al. Assuming that as many as half of the atoms are fluctuating between these two valence states, then a two-level system (TLS) would capture this configurational disorder. However, no sign of magnetism has been observed at low temperatures. Instead, we speculate that crystal-electric field effects or the freezing of self-irradiation induced Frenkel pairs at low temperatures give rise to the observed two-level system behavior. Note that nominal $^{242}Pu$ has about two orders of magnitude higher admixture of the isotope $^{238}Pu$, with a short lifetime, compared to nominal $^{239}Pu$, which is typically used in experiments.

Finally, in Fig. 11 we compare the combined total theoretical heat capacity with experiment. The agreement is excellent. Here we combined the results from Figs. 2, 8, and 10, assuming an electron-phonon coupling parameter $\lambda \approx 0.8$ with an $\alpha^2 F(\omega)$-function that has an Einstein mode at 2.8 THz, a TLS with occupation $n \approx 0.5$ and level splitting $T_{TLS} \approx 120\,\mathrm{K}$.

## V. ELECTRON-PHONON COUPLING PARAMETERS AND SUPERCONDUCTIVITY

At low temperatures ($T \to 0$) the electronic heat capacity is renormalized by the electron-phonon interaction. The enhanced Sommerfeld coefficient is $\gamma_S = \gamma_0 + \gamma_{ep} = (1+\lambda)\gamma_0$, where $\gamma_0$ is the value in the absence of the electron-phonon interaction and $\gamma_{ep}$ incorporates all these effects. We find from our low-temperature analysis of Figs. 6 through 8 a value of $\gamma_S \approx 0.0080(5)R \approx 67(4)\,\text{mJ}/(\text{mol}\,\text{K}^2)$, which is in excellent agreement with earlier estimates.[12]

The high-temperature value of the Sommerfeld coefficient, as obtained from the purely electronic contribution in Fig. 6 or Fig. 8, is not renormalized by the electron-phonon interaction and has the value of the bare electronic specific heat coefficient including all other many-electron effects, $\gamma_S \approx 0.0022(2)R \approx 18(2)\,\text{mJ}/(\text{mol}\,\text{K}^2)$. This value is remarkably close to the Sommerfeld coefficient at low temperatures in the monoclinic $\alpha$ phase of pure plutonium, $\gamma_S \approx 17-22\,\text{mJ}/(\text{mol}\,\text{K}^2)$,[6,53] suggestive that a few milli electron-volts above the Fermi level the same itinerant f-electrons are contributing in both crystallographic phases.

So far our study has revealed the presence of strong electron-phonon coupling in $Pu_{0.95}Al_{0.05}$ with a dimensionless coupling parameter of order unity, $0.8 \lesssim \lambda < 2.6$. Most cubic metals with such large electron-phonon interaction become superconducting at a few Kelvin. It is thus an intriguing question to explore at what temperature $Pu_{0.95}Al_{0.05}$ might become superconducting. The widely used McMillan's formula for estimating the transition temperature $T_c$ of a superconductor is

$$T_c = \frac{\Theta_D}{1.25} \exp\left(-\frac{1.04(1+\lambda)}{\lambda - \mu^*(1+0.62\lambda)}\right). \qquad (14)$$

where the fit parameter $\mu^*$ describes the effective Coulomb repulsion. Despite the short-comings of this expression,[21,54] and that $\mu^*$ is not known *a priori*, it has provided qualitative insights into the electron-phonon interaction and its effects on $T_c$. If we take reasonable values for $\lambda \approx 0.8 - 1.0$, $\Theta_D \approx 116\,\text{K}$, and $\mu^* \approx 0.2 - 0.3$, then we get $T_c \approx 0.4 - 4\,\text{K}$. This estimate for the superconducting transition temperature is of similar magnitude as for other strong electron-phonon coupling superconductors (see Table II), but until today no evidence of superconductivity has been observed in the $\delta$-phase stabilized Pu-Al or Pu-Ga alloys down to about 3 K, or in fcc $PuGa_3$,[55] except for superconductivity in the $\varepsilon$-phase (bcc) stabilized alloy series $(U_{1-x}Pu_x)_{0.78}Nb_{0.22}$,[56] and the tetragonal compounds $PuCoGa_5$ and $PuRhGa_5$.[57,58]

It would be interesting to know if $Pu_{0.95}Al_{0.05}$ and its fcc-stabilized sister alloys had a superconducting ground state and are not simply metastable low-temperature states above the stable ground state of the monoclinic $\alpha$ phase.

## VI. CONCLUSIONS

We studied heat capacity measurements of $\delta$-phase stabilized $Pu_{0.95}Al_{0.05}$ and calculated the vibrational, electronic, electron-phonon, anharmonic, crystal-electric field, and structural transformation contributions. Thereby, we found several important and new aspects about this material. Among these are (1) electron-phonon coupling is strong and cannot be neglected at low temperatures; (2) a flat electronic density of states, which neglects many-electron effects, yields an unphysically large dimensionless electron-phonon coupling parameter $\lambda \sim 2.6$; (3) an electronic density of states peaked at the Fermi energy with an electron-phonon coupling parameter of order unity, $\lambda \sim 0.8$, is necessary to account for most of the electronic heat capacity; (4) a remaining residual excess entropy of order $S_{res} \sim 0.4R$ can be understood in terms of an additional internal degree of freedom, for example, crystal-electric field effects or self-irradiation induced defects at plutonium sites; (5) a structural transformation from $\delta \to \alpha'$ occurs at temperatures too high, and is too small in magnitude, to account for the low-temperature excess entropy; (6) finally, the excess low-temperature entropy is not indicative of any significant lattice anharmonicity.

Clearly, more experiments are needed to resolve the relevance and transformation path of the $\delta \to \alpha'$ transition, which occurs around 130 - 180 K. Finally, the presence of crystal-electric field effects, and the possibility of a superconducting ground state for $Pu_{0.95}Al_{0.05}$, its $\delta$-phase stabilized sister alloys and $PuGa_3$, will be a challenge for low-temperature calorimetry.

TABLE II: Dimensionless electron-phonon coupling parameters and superconducting transition temperatures for various cubic metals and mercury. Here $\mu^*$ is a dimensionless fit parameter using Eq. 14.[33,59–61]

| metal | Al | Pd | Pb | V | Nb | Hg |
|---|---|---|---|---|---|---|
| structure | fcc | fcc | fcc | bcc | bcc | rhombo. |
| $\lambda$ | 0.4 | 0.7 | 1.5 | 1.0 | 1.3 | 1.6 |
| $T_c$ (K) | 1.18 | 3.2 | 7.20 | 5.46 | 7.25 | 4.15 |
| $\Theta_D$ (K) | 428 | 272 | 105 | 400 | 272 | 72 |
| $\mu^*$ | 0.115 | 0.196 | 0.229 | 0.302 | 0.330 | 0.287 |

### Acknowledgments

We thank A. Migliori, A. C. Lawson, H. Ledbetter, and R. J. McQueeney for many conversations and correspondences, and especially thank J. L. Smith for help with the calorimetry. This work was supported by the Los Alamos National Laboratory, under the auspices of the University of California for the National Nuclear Security Agency, by the US Department of Energy under Grant no. LDRD-DR 20040844 (*Phase Transformations and Strong Anharmonicities in Plutonium*). JMW and JCL also were supported in part by LDRD-DR 20030084 (*Quasiparticles and Phase Transitions in High Magnetic Fields: Critical Tests of Our Understanding of Plutonium*).